\documentclass[prb,aps,twocolumn,floatfix]{revtex4-1}
\usepackage{graphicx}
  \usepackage{amsmath}
  \usepackage{subfigure}
\usepackage{tikz}
\usepackage{amssymb}
\usepackage{amsfonts}
\usepackage{bm}
\begin{document}

\newcommand {\be}{\begin{equation}}
\newcommand {\ee}{\end{equation}}
\newcommand {\C}{\mathbb C} 
\newcommand {\Z}{\mathbb Z} 
\newcommand {\R}{\mathbb R} 
\newcommand {\I}{\mathbb I} 
\newcommand {\F}{\mathfrak F}
\newcommand {\Y}{\mathfrak Y} 
\newcommand{\beq}{\begin{eqnarray}}
\newcommand{\eeq}{\end{eqnarray}}
\newcommand{\bk}{{\bf k}}

\date{\today}
\title{Discrete Quantum Walk on a Line  with Two Entangled Particles}
\author{Joachim Nsofini}
\email{jnsofini@uwaterloo.ca}
\affiliation{$^1$Institute for Quantum Computing, University of Waterloo,\\
$^2$ Department of Physics and Astronomy, University of Waterloo,\\ 200 University Ave. West, Waterloo, Canada ,N2L 3G1}

\begin{abstract}
A review of discrete quantum walk with two particle is given. The use of different states encountered in identical particle, and the idea of entanglement and superposition is explored to explored the interesting dynamics of two particle quantum walk. Boundary conditions can specify certain dynamics and so a survey of periodic boundary condition (circle) is presented. A simulation for a Hadamard walk  for different periods of a circle is considered and  results are compared for various periods and for a case with absorbing boundaries. 
\end{abstract}
\maketitle
 
\subsection{Introduction}
 Quantum information is represented in the form of qubits 
which are similar to classical bits but have greater information carrying and processing capacities. Thus as a result  superposition of states under quantum dynamics, which progress by exploring multiple possible paths simultaneously with the amplitudes corresponding to different  paths interfering. Some of the algorithms in quantum information has been shown to have a successful implementation using quantum walks motivating why we study quantum walk in this piece of work. Quantum walks was introduced in 1993 \cite{3} to represent a quantum version of the classical walk, and has become an active field of research in quantum information and computation \cite{4}.
The variance of quantum walk on a line grows quadratically with the number of steps (time), compared to the linear growth for the classical random walk \cite{10}. A probabilistic  result is obtained in quantum walk upon measurement. Quantum walk has been broadly studied in two different forms: continuous-time quantum walk  \cite{11} and discrete-time quantum walk \cite{12}.  A  Hadamard \cite{10} discrete-time quantum walk is considered in this work. A background tool is presented to capture the nature of dynamics. Results of  two particle on a circle is presented and it comparison with a straight line is made.  
 \section{Dynamics of quantum walk}
\label{dyna}

Quantum dynamics show characteristics that do not have a classical analogue but in many cases of studies just like their classical analogue, they are considered in parallel. This is a similar case of quantum walk which is widely studied in two forms: continuous-time quantum walk and discrete-time quantum walk. 
In continuous-time quantum walk, there is a  direct definition of the walk on the position space \cite{11}, whereas in  the discrete-time quantum walk, there is a necessity to introduce a quantum coin operation to define the direction in which the particle has to move \cite{10}. These two approaches look really different but the results   are often similar, but due to the coin degree of freedom, the discrete-time variant has been shown to be more powerful than the other  in  some context  \cite{14}. To match the performance of the discrete-time quantum walk, the coin degree of freedom can be introduced in the continuous-time quantum walk  \cite{15}. Discrete-time quantum walk is the point of concern for our study.

\subsection{Continuous-time quantum walk}
\label{ctqwm}

Similar to a technique usually used in physics, it is possible to define the continuous-time quantum walk, as the quantization of a classical walk. This can be achieved by introducing quantum amplitudes in place of classical probabilities . To do this it is easier to first define the continuous-time classical random walk.

The continuous-time classical random walk takes place entirely in the {\it position} space. Consider a continuous-time classical random walk on the position space $\mathcal  H_{p}$ spanned by a vertex set $V$ of a graph $G$ with edges set $E$,  $G=(V, E)$. An adjacency matrix is used for a step of the random walk  $A$ which transform the probability distribution over $V$, i.e.,
\begin{align}
A_{j,k} =  \begin{cases}
1  &   ~~ (j,k) \in E \\
0  &   ~~ (j,k) \notin E  
\end{cases}
\end{align}
for every pair $j, k \in V$.  The matrix  $G$ is the generator matrix ${\bf H}$ given by
\begin{align}
{\bf H}_{j,k} =  \begin{cases}
d_j \gamma   &   ~~ j =  k \\
-\gamma  &   ~~ (j,k) \in E \\
0  &  ~~ {\rm otherwise}
\end{cases}, 
\end{align}
where $d_{j}$ is the degree of the vertex $j$ and $\gamma$ is the  transition probability between  neighboring nodes at any instant of  time.
\par
Let $P_{j}(t)$ be the probability of being at vertex $j$ at time $t$ then we can take the transition on graph $G$ to be the solution of differential equation
\be
\label{ctcrw}
\frac{d}{dt} P_{j}(t) = - \sum_{k \in V}  {\bf H}_{j,k} P_{k}(t).
\ee
with solutions given by
\be
P(t) = e^{-{\bf H}t} P(0).
\ee
By using quantization technique in which the probabilities $P_{j}$ are replaced by quantum amplitudes $a_{j}(t) = \langle j | \psi(t) \rangle$ where $|j\rangle$ is spanned by the orthogonal basis of the position Hilbert space $\mathcal  H_{p}$ and introducing a factor of $i$ we obtain
\be
\label{ctqw}
i\frac{d}{dt} a_{j} (t) = \sum_{k \in V} {\bf H}_{j,k} a_{k}(t).
\ee
We can see that (\ref{ctqw}) is the Schr\"odinger equation
\be
i \frac{d}{dt} |\psi\rangle = {\bf H} |\psi \rangle.
\ee
Since generator matrix is a Hermitian operator, the normalization is preserved during the dynamics. 
The solution of the differential equation can be written in the form
\be
|\psi(t) \rangle = e^{-i{\bf H}t} |\psi(0)\rangle.
\ee
Therefore, the continuous-time quantum walk is of the form of Schr\"odinger equation, a non-relativistic quantum evolution.
\par
In the  implementation of  continuous-time quantum walk  on a line, the position Hilbert space $\mathcal H_{p}$ can be written as a state span$\{  |\psi_{j} \rangle \}$, where $j \in \mathbb{Z}$. The Hamiltonian ${\bf H}$ is defined such that,
\be
\label{ctqw1b}     {\bf H}|\psi_{j}\rangle     =     -\gamma |\psi_{j-1}\rangle     +
2 \gamma |\psi_{j}\rangle  - \gamma |\psi_{j+1}\rangle  
\ee 
and  is made  to evolve with time $t$ by applying the transformation
\be
\label{ctqw1a}
 U(t) =\exp(-i{\bf H}t),
 \ee
 for time independent Hamiltonian ${\bf H}$ , and 
 \be
\label{ctqw1a}
 U(t) =\exp(-i\int{\bf H}dt),
 \ee
 for the case where the Hamiltonian is time dependent.
The Hamiltonian ${\bf H}$ of the process acts as the generator matrix which will transform the probability amplitude at the rate of $\gamma$ to the neighboring sites, where $\gamma$ is time-independent constant.

\subsection{Discrete-time quantum walk}
\label{dtqw}

We will first define the structure of the discrete-time classical random walk. The discrete-time classical random walk takes place on the position Hilbert space $\mathcal H_{p}$  with instruction from the coin operation. The coin could be fair or biased. A coin flip is used to define the direction in which the particle moves and a shift operation in succession to moves the particle in position space depending on the outcome of the coin. A simple walk on a line,  a fair coin with {\it head} and {\it tail}  can define the movement to the {\it left} and {\it right} respectively. 
\par
The discrete-time quantum walk also has a very similar structure to that of its classical counterpart.
The relevant degrees of freedom are the particle's (walker) position $i$ (with $i \! \in \! \mathbb{Z}$) on the
line, as well as its \emph{coin} state. We can then define the elements of a discrete quantum walk as follows:
\par 
 {\bf Particle or walker:} A quantum system living in a Hilbert space of infinite but
countable dimension $\mathcal H_{p}$ which is spanned by $\{|i\rangle\}$ . The walker will be initialized at the origin without lost of generality.
\par
{\bf Coin:} A quantum system living in a 2-dimensional Hilbert space $\mathcal H_{c}$ with span span\{ $|\uparrow \rangle$, $|\downarrow\rangle$ \} . This will be the quantum equivalent of randomly choosing which way the particle
will move (like tossing a coin in the classical case). The initial coin state will depends on the symmetry we want to imprint on the position probability distribution of the walker.
\par 
The total Hilbert space is given by $\mathcal H\equiv \mathcal H_{p} \otimes \mathcal H_{c}$ 
\par 
{\bf Coin Evolution Operator: } Any 2-dimensional unitary operator can be a coin evolution operator. The Hadamard operator is customary used in this work.
where $\hat{U}_C$ is the Hadamard
operator $\hat{H}$:
\begin{equation}
\hat{H}=\frac{1}{\sqrt{2}} \left[
\begin{array}{cc}
1 & 1 \\ 1 & -1
\end{array}
\right]=\frac{1}{\sqrt{2}}{\left(|0\rangle\langle 0|+|0\rangle\langle 1|+|1\rangle\langle 0|-|1\rangle\langle 1| \right) }
\end{equation}
\par 
Our system dynamics can be described by two steps, first, the \emph{coin operation}, given by $\hat{U}_C
\! \in \! SU(2)$ acting only on $\mathcal H_{c}$ and  the \emph{shift-position operation} $\hat{S}$  which will move the particle accordingly to the state of the coin. There is a possibility of quantum superposition in these operators offering a phenomenon without a classical analogue.
\par
%
\par
{\bf Conditional Shift Operator :} $\hat{S}$ will move the particle accordingly, transferring this way the quantum superposition to the total state in $\mathcal H$. As with the previous operator, the only requirement is that of unitarity. A suitable conditional shift operator is given by:
\begin{equation}\label{evol}
\hat{S} =  \left( \sum_i |i+1 \rangle \langle i | \right) \otimes
|\uparrow \rangle \langle \uparrow | + \left( \sum_i |i-1\rangle
\langle i | \right) \otimes |  \downarrow \rangle \langle \downarrow |,
\end{equation}
 The evolution of the system at each step of the walk can then be described by the total unitary operator:
\begin{equation}
\label{Eq.U} \hat{U} \equiv \hat{S} (\hat{I}_P \otimes
\hat{U}_C),
\end{equation}
where $\hat{I}_P$ is the identity operator on $\mathcal H_{p}$ and $\hat{S}$ and $\hat{U}_C$ are defined accordingly above. Note that if a measurement is performed after each step, we will revert to the classical random walk.
\par 
{\bf Observables:} Observables are defined in a standard way. In a standard way we simply mean the quantum mechanical form in which the action of an operator on its eigenstate  gives an eigenvalue and they are what we call observables. We can either measure coin states first followed by position states or measure position states directly. Results are the same because the operators commute. In the quantum  walk we will of course not measure the coin register during intermediate iterations, but rather keep the quantum correlations between different positions and let them interfere in subsequent steps. The results of interference is what gives quantum walks its powerful behaviour over the classical case.  
\par 
Consider a particle with coin degree of freedom represented in the spin. Consider a wave function representation of a spin-1/2 particle. The state is a 2-D vector $|\Psi\rangle=(|{\psi}^{\uparrow}\rangle,|{\psi}^{\downarrow}\rangle)^T$, where the first part is the component of the wave-function of the particle in the spin- $|\uparrow\rangle$ state and the second one is the component in the spin-$|\downarrow\rangle$ state. 
The composite nature of the space is emphasized by using a  tensor operator to tensor the space of the particle  and the state is given by  $|\Psi \rangle = \alpha^\uparrow |\psi^\uparrow \rangle \otimes |\uparrow\rangle + \alpha^\downarrow |\psi^\downarrow\rangle\otimes |\downarrow\rangle $, where we normalize the two wave-functions $\langle \psi^\uparrow|\psi^\uparrow\rangle=\langle \psi^\downarrow|\psi^\downarrow\rangle=1$, so that $|\alpha^\uparrow|^2+|\alpha^\downarrow|^2=1$. The tensor product '$\otimes$' separates the two degrees of freedom, spin and space, and will allow us to view the resulting correlations between these two degrees of freedom more clearly. The time development corresponding to a translation by $l$ on the larger state-space of the spin-$\frac{1}{2}$ particle can now be described by the unitary operator $U$.
This operator induces a kind of conditional translation of the particle depending on its internal spin-degree of freedom. In particular consider a situation where the spin of the particle is initially in the state $|\uparrow\rangle$, so that its wave-function is of the form $|\psi^{\uparrow}_{x_0}\rangle\otimes|\uparrow\rangle $, then application of $U$ transforms it to   $|\psi^{\uparrow}_{x_0-l}\rangle\otimes|\uparrow\rangle  $ and the particle will be shifted to the right by $l$. If the spin of the particle is in the state $|\downarrow\rangle$, and so the total wave-function is given by $|\psi^{\downarrow}_{x_0}\rangle\otimes|\uparrow\rangle $, then the translation operator will transform it to $|\psi^{\downarrow}_{x_0+l}\rangle\otimes|\uparrow\rangle $ and the particle will be shifted to the left. Quantum effects can be introduce and we have situations where the particle spin state is a cohenrent superposition of the two states. We can thus have an intial state of the form 
\be \label{Eq:init}
|\Psi_{in}\rangle=|\psi_{x_0}\rangle \otimes (\alpha^\uparrow |\uparrow\rangle + \alpha^\downarrow |\downarrow\rangle) . 
\ee
Application of the translation operator $U$ will induce a superposition of positions
\be
U|\Psi_{in}\rangle = \alpha^\uparrow  |\psi_{x_0-l}\rangle\otimes|\uparrow\rangle  + \alpha^\downarrow|\psi_{x_0+l}\rangle \otimes|\downarrow\rangle .
\ee
In this work we have used a notations that are much related to the computational states. So in order to ensure a similar approach like in the simulation, we have used $l=1$ and the position state of the particle is considered as $|i\rangle\in \mathbb{Z}$. We would equally consider a situation only where the particle start in the porition $i=0$ and the coin state is state is a coherent superposition state of $({|\uparrow \rangle + |\downarrow\rangle })/{\sqrt{2}} $
\subsubsection{Hadamard  walk}
\label{hw}

Many forms of a coin operator can be used to but we consider discrete-time quantum walk is the walk using Hadamard operation as quantum coin operation and is known as the Hadamard walk \cite{10}. A particle at origin  in one of the basis state $|0\rangle$ or $|1\rangle$ of $\mathcal H_{c}$ (internal state of the particle) is evolved into the superposition of $\mathcal H_{c}$ with equal probability, by applying the Hadamard operation
\be
H = \frac{1}{\sqrt{2}}{\left(|0\rangle\langle 0|+|0\rangle\langle 1|+|1\rangle\langle 0|-|1\rangle\langle 1| \right) }
\ee
such that
\begin{eqnarray}
\label{eq:shift}
(\hat{I}_P\otimes\hat{U}_C)  (|\psi_{0}\rangle\otimes|\uparrow\rangle) =\frac{1}
{\sqrt 2}|\psi_{0}\rangle\otimes\left ( |\uparrow\rangle+|\downarrow\rangle \right ) \nonumber \\
(\hat{I}_P\otimes \hat{U}_C) (|\psi_{0}\rangle\otimes|\downarrow\rangle) =\frac{1}
{\sqrt 2}|\psi_{0}\rangle\otimes\left (|\uparrow\rangle - |\downarrow\rangle \right ).
\end{eqnarray}
The operation $H$ is then followed by the conditional shift operation $S$ in a general form, $\hat{U} \equiv \hat{S} (\hat{I}_P \otimes\hat{U}_C))$. The  evolution of the system is allowed to progress without intermediate measurements to evolve the particle in superposition of position space and realize a  large number of steps  of the Hadamard walk. After the first  two iterations of $U$, the left and right moving components of the amplitude begin to interfere, deviating from the classical evolution, thus resulting in a quadratic speedup in the growth of the variance. The probability  amplitude distribution arising from the iterated  application  of  $U$  is significantly  different  from  the probability distribution of the classical random walk, as shown below. For the first two application, there is no difference between the classical and the quantum case, but a difference is visible as the evolution gets to more than 3 steps. The particle initially in the state $|0\rangle$ drifts to the left and the particle with an initial state $|1\rangle$ drifts to the right as explained above using notion of state vectors. This asymmetry arises from the fact that the Hadamard operation treats the two states $|\uparrow\rangle$ and $|\downarrow\rangle$ differently,  phase difference of $-1$ in case of state $|\downarrow\rangle$.  This phase difference, brings our some of the significant differences depending on the initial state of the particle contributes to the constructive interference on one side and to the destructive interference on the other side of the position space.
%
 Therefore, if the initial state of our particle is, for instance
$|0 \rangle \otimes | \uparrow \rangle$, the first step of the
quantum walk will be as follows:
\begin{eqnarray}
|0 \rangle \otimes | \uparrow \rangle
&\stackrel{\hat{U}}{\longrightarrow}& \frac{1}{\sqrt{2}} \left(
|1\rangle \otimes | \uparrow\rangle + |-1\rangle \otimes |
\downarrow\rangle \right).\nonumber \\
&\stackrel{\hat{U}}{\longrightarrow}& \frac{1}{2} \left(
|2\rangle \otimes | \uparrow\rangle -|0\rangle \otimes (| \uparrow\rangle - 
\downarrow\rangle )+ |-2\rangle \otimes |
\downarrow\rangle \right).\nonumber \\
&\stackrel{\hat{U}}{\longrightarrow}& \frac{1}{2\sqrt{2}} \left(
|3\rangle \otimes | \uparrow\rangle + |1\rangle \otimes | \downarrow\rangle +|-1\rangle \otimes | \uparrow\rangle -2|-1\rangle \otimes| \downarrow\rangle - |-3\rangle \otimes |
\downarrow\rangle \right).\nonumber \\
\end{eqnarray}
This illustrates that there is a similar probability  to find the particle in position 0,1 and 2 for classical and quantum walks for zeroth, first and second steps. However there is a striking difference for the third and higher order, situation explained only by superposition principle, and this is the effect that is explored in most quantum application. Applying the coin unitary operator without intermediate measurements we can then obtain a plots shown below for below. The behaviour depends on the nature of the initial state. When the coin initial state is not a coherent superposition, we obtain an asymmetric situation given by:
\begin{figure}[h]
\begin{center}
\includegraphics[scale=.3]{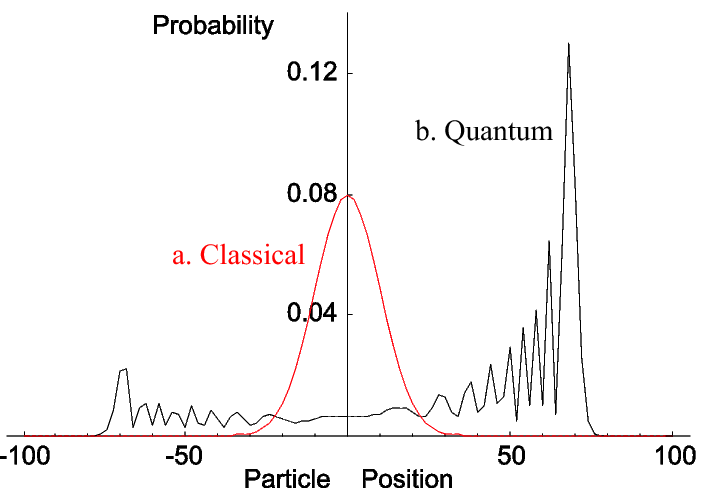}
\caption{Probability distribution of classical random walk (Red line Gaussian) and its quantum counterpart, quantum walk after $N=100$ steps \cite{2}. The initial state is $|0 \rangle \otimes |\uparrow \rangle$.   } 
\label{cqpd} 
\end{center}
\end{figure}
\begin{figure}[h]
\begin{center}
\includegraphics[scale=.3]{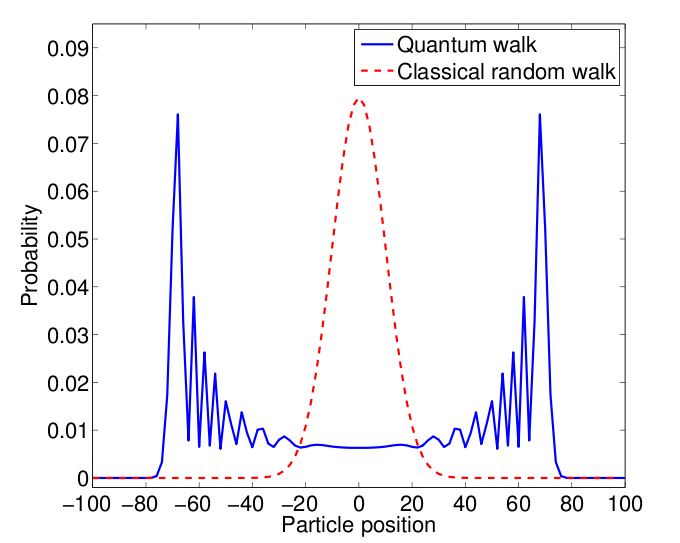}
\caption{Probability distribution of classical random walk (Red line Gaussian) and quantum walk after each after $N=100$ steps \cite{16}.The initial state is $\frac{1}{2}|0 \rangle \otimes(|\uparrow \rangle + |\downarrow \rangle$.   } 
\label{cqpd2} 
\end{center}
\end{figure}
The plot in figure \ref{cqpd} shows a plot for a quantum and a classical case after, $N=100$ steps. The difference in the variance can been seen in the distribution. The quantum case is shown for the non symmetric state $|0 \rangle \otimes |\uparrow \rangle$ . The asymmetry can be uplifted by by choosing the coin in a coherent superposition of the |$\uparrow \rangle $ and $ |\downarrow \rangle$ as shown below.

The plot in figure \ref{cqpd2} shows a plot for the classical and quantum walk after $N=100$ time steps. Here the curve is symmetric because the coin start in a coherence superposition  state.
%
\subsubsection{Discrete-time quantum walk on a periodic line}
\label{dtqwc}
The quantum walk on a particle in a periodic line can be simulated using an n-cycle. This can be viewed as a particle moving in a circle with nodes. This walk will then depend on the number or nodes in a circle or in other words the periodicity of the circle. Before introducing the concept of quantum limiting distribution, we provide an example of a
quantum walk on a periodic graph: a discrete quantum walk on a cycle.
%
A cycle is a graph $G$ with vertex set $V$,($|V=n|$) and  edges set $E$ ,  $G=(V, E)$. A quantum walk on $G$ acts on a total Hilbert space $\mathcal H_{p} \otimes \mathcal H_{c}$  . In this case of the Hadamard coin operator given above the shift operator is defined by $\hat{S}|0,j\rangle=|0,j+1 $ mod $ n\rangle$ and $\hat{S}|1,j\rangle=|0,j-1 $ mod $ n\rangle$. The behaviour of quantum walk on a cycle as will be shown depends so strongly on the the number of nodes or periodicity of the cycle. Here we shall be concern with a circle only which is a cycle with $|V|=2$. Below is a a diagram showing quantum walk for a single particle on a circle with different number of edges or period $T$.
\begin{figure}[h]
\begin{center}
\includegraphics[scale=.4]{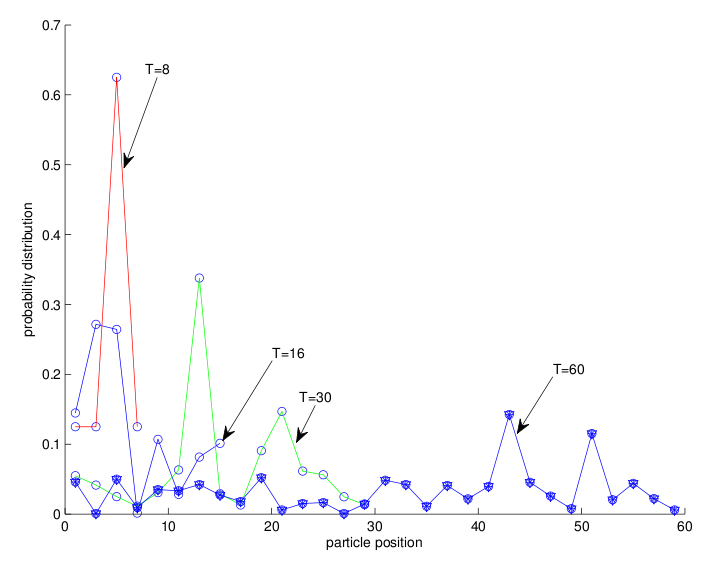}
\caption{Quantum walk on different periodic cycle for a single walker. The probability distribution depends so much on the number of nodes and its shown for different number of nodes $T$. 
 }
\label{quantumwalk} 
\end{center}
\end{figure}
\section{Quantum Walk of Two Particle}
 
\subsection{Introduction}
 
Quantum walk has been shown to have features with no classical analogue \cite{5,6,7,8} . Just like in the previous section we will consider quantum walk in system with two particles. We consider two unitary operators to represent our quantum equivalence of a coin and then we have two walkers. There are many features that  can be explored  including the notion of superposition and entangled states \cite{2} for which no known classical equivalence exist. Entanglement can be studied in the coin degree of freedom as well as in the particle degree of freedom. In this section we will present studies made only in the coin degree of freedom which we use an analogy of two particles with spin states that are entangled. Again we will present results that suggests that the behaviour of these walkers is also a function of the type of graph they traverse as we will compare and contrast evolution on a graph with and without periodic  boundary conditions. In this section we will start by explaining quantum walk of two particle in a line and then we extend it to a cycle and make some comparisons to the results. 
\subsection{Dynamics of two particle walk}
Two particles in a system can be interacting or non interacting. In the case of quantum walk, such a dynamic will live in a composite Hilbert space denoted by
\begin{align}
 \mathcal H_{12}\equiv \mathcal H_{1} \otimes \mathcal H_{2}\equiv \mathcal H_{p,1} \otimes \mathcal H_{c,1} \otimes \mathcal H_{p,2} \otimes \mathcal H_{c,2} 
 \end{align}  
where $\mathcal H_{1}$ and $\mathcal H_{2}$ represent the Hilbert spaces of particles $1$ and $2$ respectively. Since the relevant degrees of freedom in our problem are the same for both particles, we have that both $\mathcal H_{1}$ and $\mathcal H_{2}$ are isomorphic to $\mathcal H$ defined earlier for the one-particle case. 
The dynamics of of such a system is govern by a unitary given by
\begin{equation}
\hat{U}_{12}= \hat{U} \otimes \hat{U},
\end{equation}
where $\hat{U}$ is given by equation (\ref{Eq.U}) and is the same for both particles. Note also that in the case of identical particles we have to restrict $\mathcal H_{12}$ to its symmetrical and antisymmetric subspaces, respectively for bosons and fermions. The other entangled states of the triplet states are not considered as all the of them are symmetric and gives the same results when considered. Let us then consider without lost of generality the case where both particles start the quantum walk in the same position state $|0,0\rangle$, but with different coin states $|\downarrow\rangle$ and $| \uparrow\rangle$. Consider a simple way to understand this mathematically. Let us consider the coin subspace by using a representation with $|0\rangle\equiv| \uparrow\rangle$ and $|1\rangle\equiv| \downarrow\rangle$. The coins consists of two qubits with basis $\{ |j,k\rangle\} ,j,k\in\{0,1\} $. The particle position space is given by $\{ |m,n\rangle\} ,m,n\in\mathbb{Z} $ ,and then the generic state of the quantum walker at any time step $t$ would be
\begin{equation}\label{E:psi}
|\psi(t)\rangle=\sum_{j,k}\sum_{m,n}A_{j,k;m,n}(t)|j,k\rangle|m,n\rangle.
\end{equation}
We can also represent the evolution operator $ \hat{U} \equiv \hat{S} (\hat{I}_P \otimes \hat{U}_C),$ in this notation where, 
\begin{equation}
\label{Eq.CC} 
C=\sum_{j,k}\sum_{j^{\prime},k^{\prime}}C_{j,k;j^{\prime},k^{\prime}}|j,k\rangle \langle j^{\prime}k^{\prime}|
\end{equation}
is the coin operator, $I_P$ is the identity matrix, and $S$ is the shift operator given by equation \ref{evol} and which we represent here by equation \ref{Eq.SS}
\begin{equation}
\label{Eq.SS} 
S|j,k\rangle|m,n\rangle =|j,k\rangle|m + (-1)^j,n + (-1)^k\rangle.
\end{equation}
We can see that the two particles both moves to the right when the coin is $|0,0\rangle$, and to the left when the coin is $|1,1\rangle$. In the state $|1,0\rangle$, particle one moves to the left and particle two moves to the right, and the situation is reverse for the state $|0,1\rangle$. 

Applying the evolution operator on state \ref{E:psi} we get
\begin{equation}
\label{Eq.CA} 
A_{j,k;m,n}(t+1)=\sum_{j^{\prime},k^{\prime}}C_{j,k;j^{\prime},k^{\prime}}A_{j^{\prime},k^{\prime};m + (-1)^j,n + (-1)^k}(t).
\end{equation}
The probability distribution for the walker at position $|m,n\rangle$ at time t is
\begin{equation}
\label{Eq.P} 
P_{m,n}(t)=\sum_{j,k}|A_{j,k;m,n}(t)|^2.
\end{equation}

We can then go back to our normal formalism to represent the possible states as is done below.
There will be three cases we shall consider here. The  pure separable state where  our system's initial state will be given by:
\begin{equation}
\label{Eq. Initial-S}
|\psi_0^S\rangle_{12}=|0,\downarrow\rangle_1|0,\uparrow\rangle_2.
\end{equation}
Without any confusion, we have switched back to the normal notation where the $'0'$ in the state above is showing the particle's position and has nothing to do with the coin degree of freedom.
The other situation that is considered is for initial pure entangled states in the coin degree of freedom. The case  of two maximally entangled coins is considered. The symmetric case where  our system's initial state will be given by:
\begin{equation}
\label{Eq. Initial-E}
|\psi_0^{+}\rangle_{12}=\frac{1}{\sqrt{2}}(|0,\downarrow\rangle_1|0,\uparrow\rangle_2
+|0,\uparrow\rangle_1|0,\downarrow\rangle_2),
\end{equation}
and the antisymmetric case where  our system's initial state will be given by:
\begin{equation}
\label{Eq. Initial-EE}
|\psi_0^{-}\rangle_{12}=\frac{1}{\sqrt{2}}(|0,\downarrow\rangle_1|0,\uparrow\rangle_2
-|0,\uparrow\rangle_1|0,\downarrow\rangle_2).
\end{equation}
%
%
These states differ only by the presence of a phase. These are states that are considered as bosonic and fermion particles in physics when the notion of identical particles is introduced. After an evolution of $N$ time steps a system starting with initial state $|\psi_0\rangle_{12}$ becomes;
\begin{equation}
|\psi_N^S\rangle_{12}=\hat{U}^N_{12} \, |\psi_0^S\rangle_{12}=
\hat{U}^N |0,\downarrow\rangle_1 \hat{U}^N |0,\uparrow\rangle_2.
\end{equation}
 for initial conditions (\ref{Eq. Initial-S}). The probability distribution $P_{12}^S (i,j;N)$ for finding particle $1$ in position $i$ and particle $2$ in position $j$ for $N=100$ steps is shown in section (\ref{sec3}). We can show that the distributions of the particles are uncorrelated, and so $P_{12}^S (i,j;N)$ is simply the product of the two independent one-particle distributions:
\begin{equation}
P_{12}^S (i,j;N)=P^S_1(i;N) \times P^S_2(j;N),
\end{equation}
where $P^S_1(i;N)$ is the probability distribution for finding
particle $1$ in position $i$ after $N$ steps, and similarly 
$P^S_2(j;N)$ for particle $2$.
A corresponding results for movement on a cycle is given beside.
\par 
In a similar fashion, when the particles are entangled, evolution of the system after $N$ steps will be 
\begin{equation}
\label{Eq. Initial-Ev1}
|\psi_N^{+}\rangle_{12} = \hat{U}^N_{12} \,|\psi_0^{+}\rangle_{12}
                        =\frac{1}{\sqrt{2}}(\hat{U}^N|0,\downarrow\rangle_1\hat{U}^N|0,\uparrow\rangle_2
+\hat{U}^N|0,\uparrow\rangle_1\hat{U}^N|0,\downarrow\rangle_2),
\end{equation}
for the symmetric entangled state, and 
\begin{equation}
\label{Eq. Initial-Ev2}
|\psi_N^{-}\rangle_{12} = \hat{U}^N_{12} \,|\psi_0^{-}\rangle_{12}
                        =\frac{1}{\sqrt{2}}(\hat{U}^N|0,\downarrow\rangle_1\hat{U}^N|0,\uparrow\rangle_2
-\hat{U}^N|0,\uparrow\rangle_1\hat{U}^N|0,\downarrow\rangle_2),
\end{equation}
for the antisymmetric case.

The figures show the probability distribution for finding particle $1$ in position
$i$ and particle $2$ in position $j$ in the "+" case, $P_{12}^+ (i,j;N)$,  and $P_{12}^- (i,j;N)$ of the "-" case for $N=100$. The effects of the entanglement are striking as noticed when comparing them as shown in the distributions plotted in section ( \ref{sec3}) . There is a  significant increase the probability of finding the particles in certain configurations and some are forbidden. This is one reason why there is a speed up in quantum algorithms based on quantum walk.
The results for particle dynamics on a circle is also presented. There are some similarities as well as some striking differences. We can note from the graphs that the particles tend to finish as far as possible in the case of antisymmetric states a situation much common to nature and also a situation that doesn't manifest itself well in periodic conditions, except for large enough periods compared to the total time time steps. To make a concrete comparison we will compare the dynamics of each particle for both the walk on a line and on a circle by considering their average positions and comparing for the three different states for both line and circle. We will also study the variations in their Von Neumann entropy for a general configuration and also for some specific positions.
\section{Comparison of Results}\label{sec3}
\subsection{Introduction}
The discrete quantum walk of two particle on a straight line, shows interesting behaviours depending on the boundary condition. As shown in \cite{2} there are many interesting behaviour that are a result of the initial state. The underlying dynamics of each particle is also presented, but in this chapter we shall compare results for various boundary conditions. We present results of quantum walk on a line for product, symmetric and antisymmetric coin state.

\subsection{Absorbing boundary conditions.}
The following results illustrate the difference obtained for different initial states. The results shows some striking difference. This is as a result of some nontrivial quantum correlations resulting from the measurements in the case of symmetric and antisymmetric states. The results is given below,

\begin{figure}[h]
\begin{center}
\includegraphics[scale=.3]{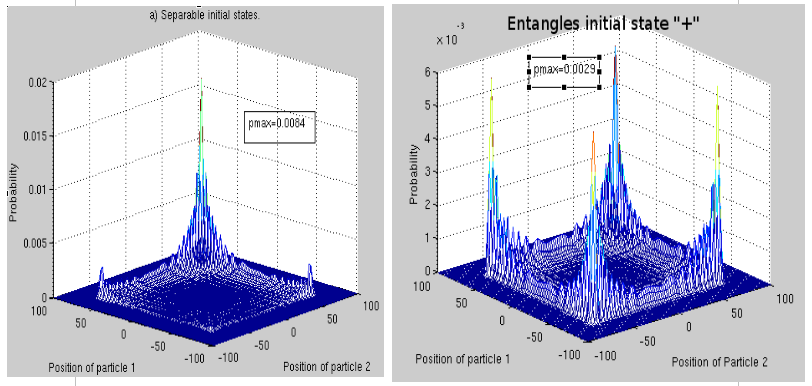}
\caption{The probability distribution for two particles on a line after $N=100$ time steps. The result is shown for product and symmetric states.
 }
\end{center}
\end{figure}
Also in a similar way the results for product and antisymmetric states is shown below.
\begin{figure}[h]
\begin{center}
\includegraphics[scale=.3]{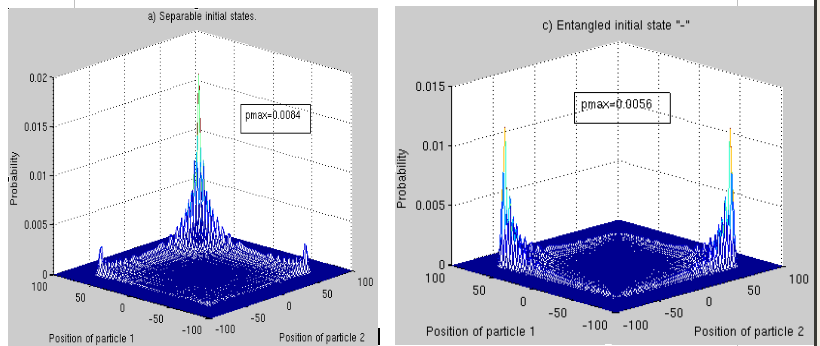}
\caption{The probability distribution for two particles on a line after $N=100$ time steps. The result is shown for product and symmetric states.
 }
\end{center}
\end{figure}
These behaviors are investigated in \cite{2} , in which they studied qualitatively the behavior of each of the particle using partial trace knowledge to get a reduced density operator for the evolution of each particle. Using our knowledge of bosons and fermions, we see that in the case of identical particles, in the symmetric state (bosons), the particles tend to finish together, while in the antisymmetric state (fermion), the particles tend to finish as far apart as possible, offering a behaviour we expect, from their statistics.

\subsection{Periodic boundary conditions}
Using the results above a survey for the various initial states, and then a implementation of the boundary conditions of a circle gives results that depends on the boundary conditions. A study made with two particles and for different types of boundary conditions is made. The probability distribution is periodic for smaller periods. It can also be noticed from the distribution that some regions on the circle are favored. A comparison between periodic and absorbing boundary conditions is made below. 
\begin{figure}
\centering
\includegraphics[scale=.3]{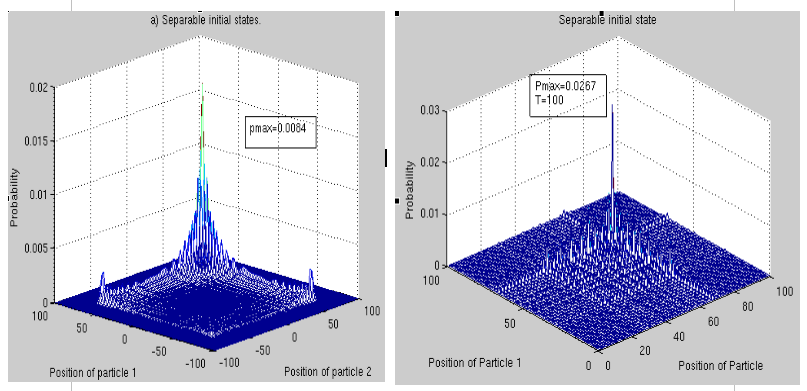} \\
\includegraphics[scale=.3]{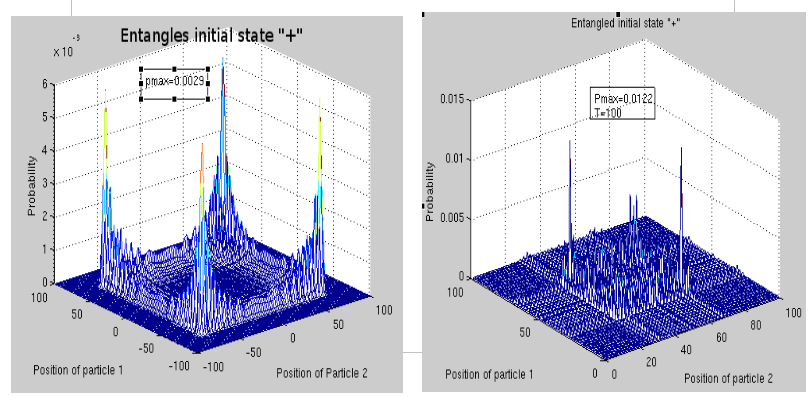} \\
\includegraphics[scale=.3]{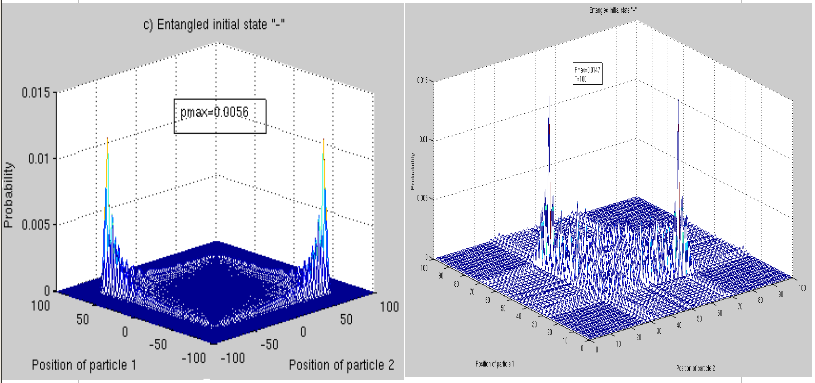} 

\caption{The probability distribution for two particles on a line after $N=100$ time steps. The result is shown for product, symmetric and antisymmetric states in absorbing and non  absorbing boundary conditions. Here a Period $T=100$ is used to ensure enough compatibility.
 } 
\end{figure}
There are different ways to study the behaviour of each of the situation. A reduced density could be used as already pointed above. One way to characterize the behaviour of the three systems is to study their Von Neumann entropy as already pointed above. With the entropy of the combined system in mind, we can get a reduced description as well. With this we can study the behaviour of a single particle in the composite system. Various studies can then be made of the probability distribution for seeing the particle at position, $i$ after $k$ time steps.So fixing $i$ the position we can obtain a plot for the probability distribution for seeing the particle at that position after each time step.

\subsection{Different periods}

\begin{figure}\label{ffigss}
\centering
\includegraphics[scale=.3]{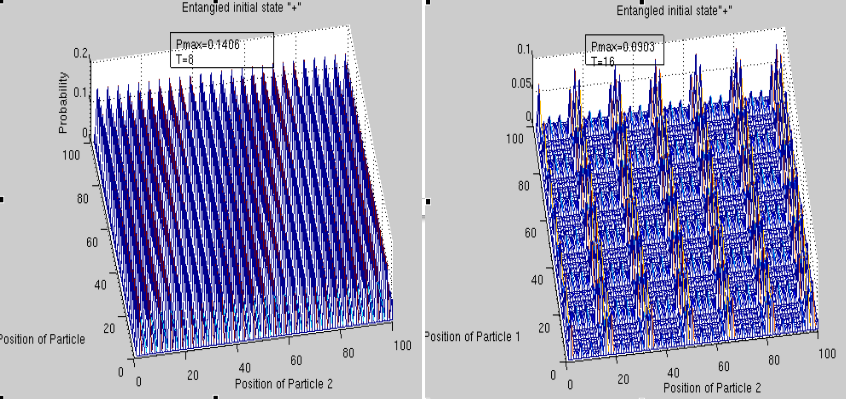} 
\includegraphics[scale=.3]{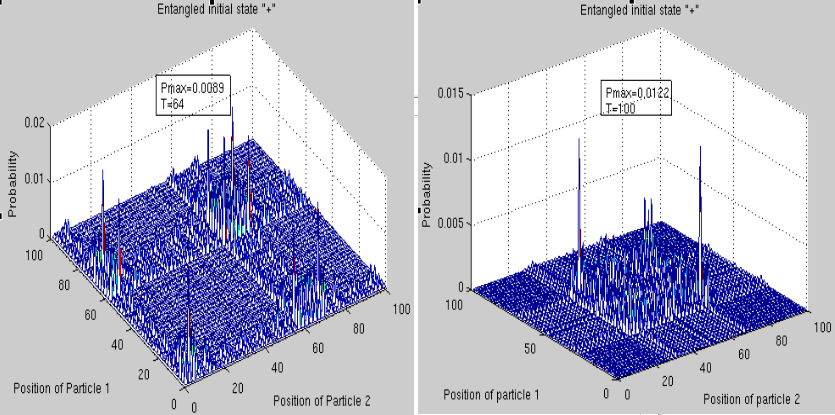} 
\caption{The probability distribution for two particles on a line after $N=100$ time steps. The result is shown for product and antisymmetric states in absorbing and non  absorbing boundary conditions. Here a total time  $T=100$ is used to make comparison for different periods.
 } 
\end{figure}

A study of the period shows that the distribution depends so much on the periodicity of the system. Below are plots obtained for the symmetric state for the periods $T=8,16,64$ and $100$. We make the plot in the form of a lattice, i.e a periodic lattice and the we can observe periodicity on the lattice as the particle undergo its dynamic.By lattice here, we mean that the particle after traversing the period, comes to a point that is identical to the original pint and so we have a possibility to see the periodicity in the plot. Another way could be to make a plot that is limited to nodes equal to the period. This is less illustrative as it doesn't explicitly show the periodic nature of the results. A study of the probability distribution to see a particle at  position, $i$ after $k$ time steps, plotted for each $i$ would can be made. A similar results for the one particle case is shown by \cite{17} in which the showed that the probability distribution for seeing the particle at the starting position is periodic up to $T=8$ and that the system become really chaotic with an inability to predict where it can be from it entropy for $T=16,32, 64$ and 100. The results for different periods is shown figure \ref{ffigss}


\section{Acknowledgements}
I offer my sincere thanks to Yasser Omar for enlightening discussions and for proposing such a great topic.
My appreciation to the  African University of Science and Technology for accommodating me during the period that I started this work.


\end{document}